\documentclass[a4paper,11pt]{article}
\pdfoutput=1 

\usepackage{jinstpub} 

\usepackage{lineno}
\usepackage{textgreek}

\title{\boldmath Ultra-low jitter clock distribution for the trigger electronics of the New Small Wheel for the ATLAS experiment}

\author[a]{T.\,Alexopoulos}
\author[b]{T.\,Geralis}
\author[c,1]{P.\,Gkountoumis\note{Corresponding author.}}
\author[d]{L.\,Levinson}
\author[e]{I.\,Mesolongitis}
\author[b]{O.\,Zormpa}

\affiliation[a]{National Technical University of Athens,\\9, Iroon Polytechniou St., Zografou, Greece}
\affiliation[b]{National Centre for Scientific Research "Demokritos",\\Patr. Gregoriou E \& 27 Neapoleos Str, Agia Paraskevi, Greece}
\affiliation[c]{CERN,\\Geneva 23, Meyrin, Switzerland}
\affiliation[d]{Weizmann Institute of Science,\\234 Herzl St., Rehovot, Israel}
\affiliation[e]{University of West Attica,\\250 Thivon \& P.\,Ralli St., Egaleo, Athens}

\emailAdd{panagiotis.gkountoumis@cern.ch}

\abstract{The Large Hadron Collider (LHC) at CERN plans to have a series of upgrades to increase its instantaneous luminosity to $7.5\times10^{34}\,\mathrm{cm^{-2}s^{-1}}$. 
The increased luminosity  drastically impacts the ATLAS trigger and readout data rates. The inner-most station of the ATLAS muon  spectrometer, the so-called Small Wheels is being replaced with a New Small Wheel (NSW) system, consisting of Micromegas and small-strip Thin Gap Chambers (sTGC) detectors. 
The on-detector radiation levels required radiation tolerant electronics.
The lower radiation levels on the rim of the NSW allowed utilizing commercial electronic chips, such as Field Programmable Gate Arrays (FPGAs), in the trigger chain of the sTGC detectors. 
Those FPGAs require an ultra-low jitter clock for the proper operation of their Gigabit transceivers ($4.8\,\mathrm{Gb/s}$ serial links). 
The initial design was based on a clock provided by a radiation tolerant ASIC designed at CERN, but due to its intrinsic jitter and consequent marginal error rate on the transmission lines, a different approach had to be chosen. 
An additional clock source based on commercial jitter cleaners, fanout chips and optical transmitters driving dedicated fibers was built.
The new scheme  provides 64 low-jitter clocks (32 main and 32 redundant) from the radiation-protected area (USA15) to the trigger electronics over $60\,\mathrm{m}$ of OM3 fiber. 
}


\arxivnumber{} 

\proceeding{
TWEPP 2021 Topical Workshop on Electronics for Particle Physics\\
  Sep 20 – 24, 2021 \\
  On-line}

\begin{document}
\maketitle
\flushbottom


\section{Introduction}
\label{sec:intro}

The sTGC trigger electronics are placed in an accessible position on the rim of the NSW about $5\,\mathrm{m}$ away from the interaction point. 
The low radiation levels at this point gave the opportunity to utilize commercial devices for the trigger electronics. 
The simulated radiation levels for the Total Ionizing Dose (TID), Non-Ionizing Energy Loss (NIEL), Single Event Effects (SEE) and magnetic field are presented in Table\,\ref{tab:rad_levels}. 
To transfer the high amount of sTGC trigger data from the on-detector Front-end boards and to perform multiple calculations on the coincidence logic, FPGAs and high bandwidth serial links at $4.8\,\mathrm{Gb/s}$ were utilized\cite{a}. 

For the serialization and data transmission from the front-end boards to the trigger electronics the Trigger Data Serializer (TDS)\cite{b} ASIC was utilized.
The later receives the $40\,\mathrm{MHz}$ LHC clock by the GigaBit Transceiver (GBTX)\cite{c} ASIC from CERN (housed on the sTGC-Level-1 Data Driver Card (L1DDC)\cite{d} board) and generates a $160\,\mathrm{MHz}$ reference clock using an internal PLL.
The output data use the Low Voltage Signalling (SLVS) standard.

\begin{table}[htbp]
\centering
\caption{\label{tab:rad_levels} Simulated Radiation Loads and Magnetic Fields, from\cite{e} for the NSW
after 10 years at high luminosity LHC.}
\smallskip
\begin{tabular}{|c|c|c|}
\hline
                        & Inner radius (R=1 m)                      & Outer Rim (R=5 m) \\
\hline
TID (\textgamma)        & $400\,\mathrm{Gy}$                        & $16\,\mathrm{Gy}$\\
NIEL (fast neutrons)    & $2.3\times{10^{13}}\,\mathrm{n/cm^{2}}$   & $7.3\times{10^{11}}\,\mathrm{n/cm^{2}}$\\
SEE (protons)           & $4.2\times{10^{13}}\,\mathrm{p/cm^{2}}$   & $1.3\times{10^{11}}\,\mathrm{p/cm^{2}}$\\
B field                 & $\leq1\,\mathrm{kG}$                      & $5\,\mathrm{kG}$\\
\hline
\end{tabular}
\end{table}

The FPGA transceivers require an extra low jitter reference clock to ensure the signal integrity of the transmitting and receiving data. 
Furthermore, due to the attenuation of the high speed signals over the long twinaxial cables (up to $6\,\mathrm{m}$) repeater chips were also utilized.
The attenuation of tin and silver cables is presented in Table\,\ref{tab:twinax_attenuation}.

\begin{table}[htbp]
\centering
\caption{\label{tab:twinax_attenuation} Twinax cable attenuation for tin and silver-plated cables.}
\smallskip
\begin{tabular}{|c|c|c|c|c|c|c|c|}
\hline
Frequency (GHz) & 0.5 & 1.0 & 2.0 & 5.0 & 10.0 & 15.0 & 20.0\\
\hline
Tin Plating (dB/m) & -0.90 & -1.4 & -2.2 & -4.0 & -7.5 & -10.9 & -14.6 \\
Silver Plating (dB/m) & -0.85 & -1.2 & -1.7 & -3.2 & -4.9 & -6.8 & -8.8 \\
\hline
\end{tabular}
\end{table}

\begin{figure}[htbp]
\centering 
\includegraphics[width=1\textwidth]{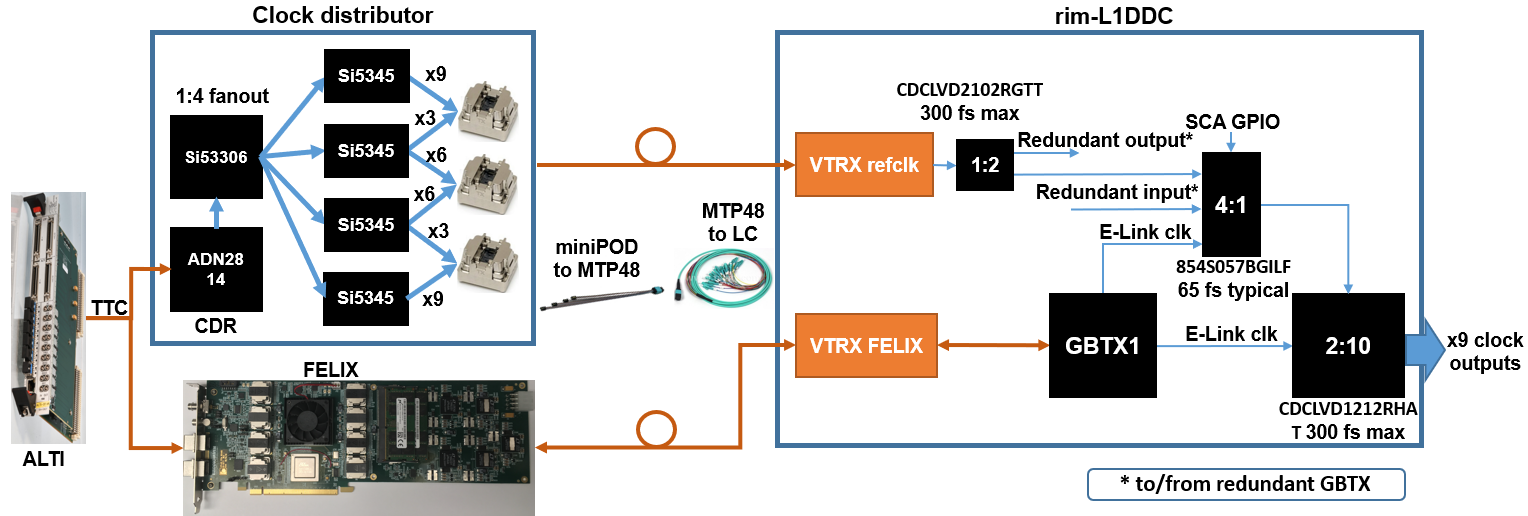}
\caption{\label{fig:overall_scheme} Clock distribution for the NSW trigger electronics.}
\end{figure}

Initially the E-Link and programmable clocks of the GBTX ASIC were used as reference clocks but they proved to be marginal during the data integrity tests due to their high jitter of about $4.3\,\mathrm{ps}$ and $10\,\mathrm{ps}$ respectively. 
For this purpose a new clock distribution scheme was designed to deliver a direct low jitter clock to the 288 trigger boards of the NSW over $63\,\mathrm{m}$ fibers, but also to preserve the E-Link clocks as a selectable option. This scheme is based on the Si5345 jitter cleaner by Silicon labs and a fanout network.
The overall scheme, presented in Figure\,\ref{fig:overall_scheme}, can be divided into two categories: the on-detector electronics Rim-L1DDC\cite{e}) comprised mainly of radiation tolerant components and the off-detector electronics (ALTI, FELIX and clock distributor board) housed in the radiation-protected area (USA15) consisting of Commercial-Off-The-Shelf (COTS) components.
The clock distributor board receives the timing, trigger and control (TTC) stream from the ALTI module and recovers the LHC clock by using a Clock and Data Recovery (CDR) chip (AD2814) from Analog Devices. 
Then, with the help of a 1:4 fanout buffer (Si53306) from Silicon Labs with an Ultra-low additive jitter of $50\,\mathrm{fs}$  the clock is fed to four Si5345 jitter cleaners. 
Each board provides 32 clocks that are distributed to the on-detector electronics via three 12-channel micro-POD optical transmitter modules and MTP fibers. 
One distributor board will be used for each wheel providing in total 64 clocks (32 main and 32 redundant).
The L1DDC board is equipped with the CERN GBTX and Slow Control Adapter Adapter (SCA) ASICs. 
The output clock can be selected to be either the GBTX E-Link or the direct clock with the fanout network. 
The fanout network consists of 1:2 and 2:12 fanout buffers from Texas Instruments with less than $300\,\mathrm{fs}$ additive jitter (part numbers CDCLVD2102RGTT and CDCLVD1212RHAT respectively) and a 4:1 fanout buffer from IDT with less than $65\,\mathrm{fs}$ additive jitter (part number 854S057BGILF).

\section{The clock distributor board}
\begin{figure}[htbp]
\centering 
\includegraphics[width=0.6\textwidth]{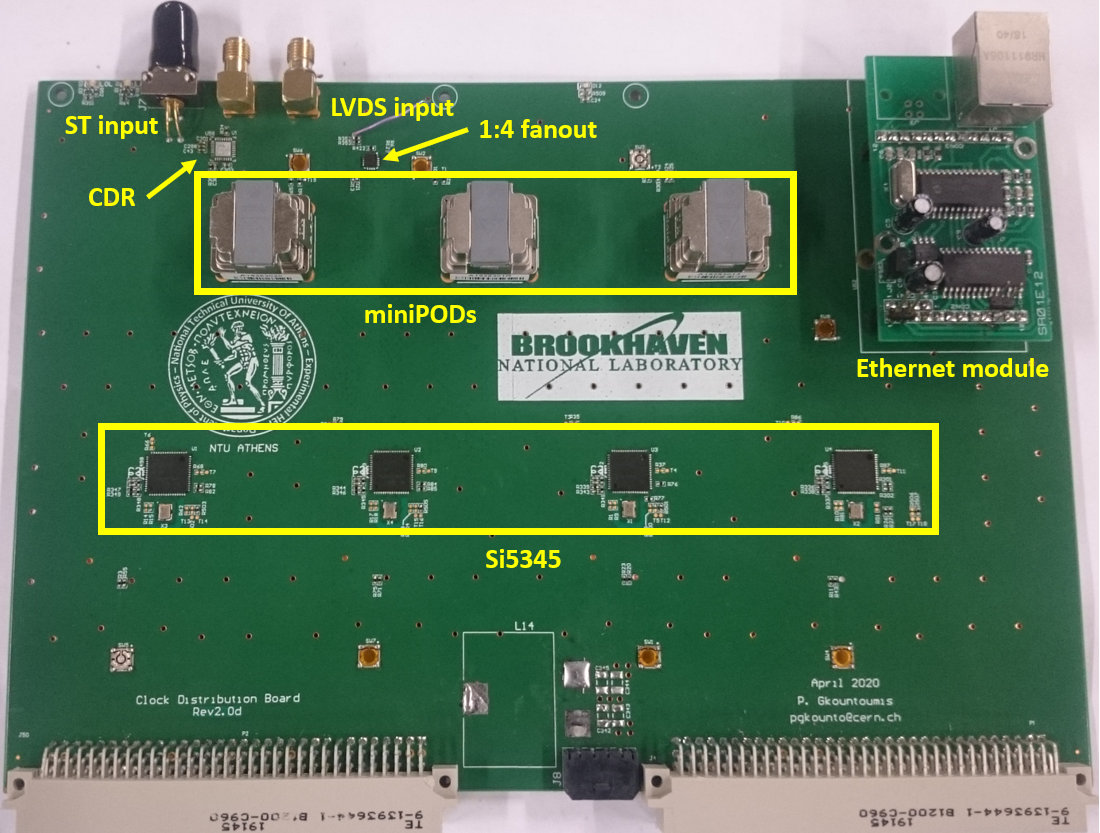}
\caption{\label{fig:distributor} The clock distributor board.}
\end{figure}

The clock distributor board is a VME\,6U blade that receives the TTC stream from ALTI module over a fiber with an ST connector. An additional LVDS input over SMA connectors bypassing the clock and data recovery chip is also available. Specific layout rules were incorporated to reduce any potential noise that could increase the output jitter. Those rules include specific component placement, shielding of the differential pairs in the internal layers, utilizing dedicated LDOs for each Si5345 channel, use of low ESR capacitors and passive filtering on the input power rail. The Si5345 and microPODs are configured via a commercial Ethernet-to-I\textsuperscript{2}C module running at $10\,\mathrm{Mb/s}$. The Ethernet connection obviates the need for a VME single board computer to configure the board. The clock distributor board is presented in Figure\,\ref{fig:distributor}.

\section{Radiation tests}

The 4:1 fanout chip of the L1DDC board was irradiated with a total dose of $6.54\times10^{10}\,\mathrm{n/cm^2}$ and maximum flux of $4.6\times{10^6}\,\mathrm{n/cm^2/s}$ at the 5.5\,MV HV TN-11 TANDEM Van De Graaﬀ accelerator of NCSR ``Demokritos'' in Athens.  
The 2:4 fanout chip with part number CDCLVD1204 which is the same family and technology (CMOS) with the 1:2 and 2:12 fanout chips of the L1DDC was also validated under radiation at Demokritos with a  $4.82\times10^{10}\mathrm{n/cm^2}$ (maximum flux of $2.82\times{10^6}\,\mathrm{n/cm^2/s}$) and $^{60}$Co facility at CERN with a TID of $670\,\mathrm{Gy}$ ($13.14\,\mathrm{Gy/H}$).
In both cases the clock generated by a synthesizer was fed to the PLL of the FPGA via the fanout chips where the lock signal was monitored and a counter based on the received clock was compared to a local one.
In all cases no errors or failures were observed during the irradiation period.

\section{Results}
The whole scheme was evaluated by using the clock distributor with a Si5345 evaluation board as a clock source over the LVDS input and total $110\,\mathrm{m}$ ($2\times 50\,\mathrm{m}$ + $60\,\mathrm{m}$ with couplers) fiber cables. 
The output of the Rim-L1DDC board was driven to a signal source analyzer with a use of a mini Serial Attached SCSI mini-SAS to SMA adapter. 
The maximum phase noise jitter was measured to be $1\,\mathrm{ps}$ with an integrated bandwidth from $1\,\mathrm{kHz}$ to $30\,\mathrm{MHz}$. 
Xilinx\textsuperscript{\textregistered} provides the phase noise mask of their 7-series GTX/GTH/GTP transceivers for  $156.25\,\mathrm{MHz}$ and therefore a correction factor to the mask point according to the equation shown below was applied. 

    $$PhaseNoiseAt(X)\mathrm{MHz}=PhaseNoPhaseNoiseAt(Y)\mathrm{MHz}+20\times{log\frac{X}{Y}}$$
    
\begin{figure}[htbp]
\centering 
\includegraphics[width=1\textwidth]{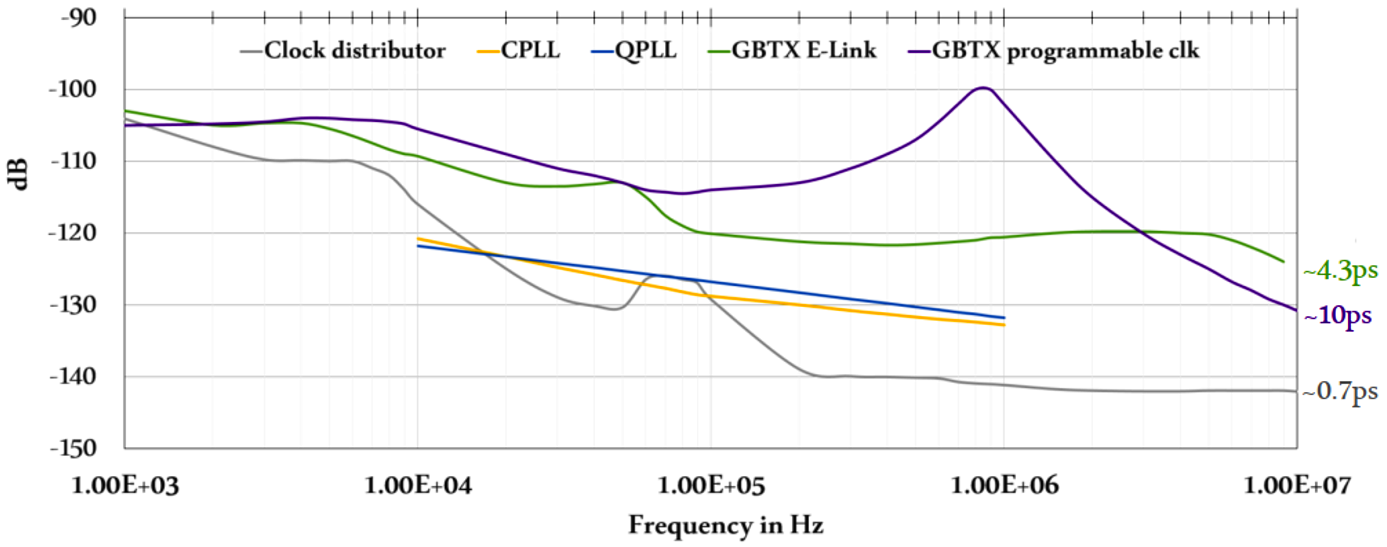}
\caption{\label{fig:phase_noise} Phase noise jitter of the direct and GBTX clocks compared to Xilinx\textsuperscript{\textregistered}'s phase noise mask.}
\end{figure}

The Phase noise jitter of the GBTX E-Link clock, the GBTX programmable-phase clock and the direct clock of the clock distributor described here compared to Xilinx\textsuperscript{\textregistered}'s phase noise mask for CPLL and QPLL are presented in Figure\,\ref{fig:phase_noise}.
   
\begin{figure}[htbp]
\centering 
\includegraphics[width=1\textwidth]{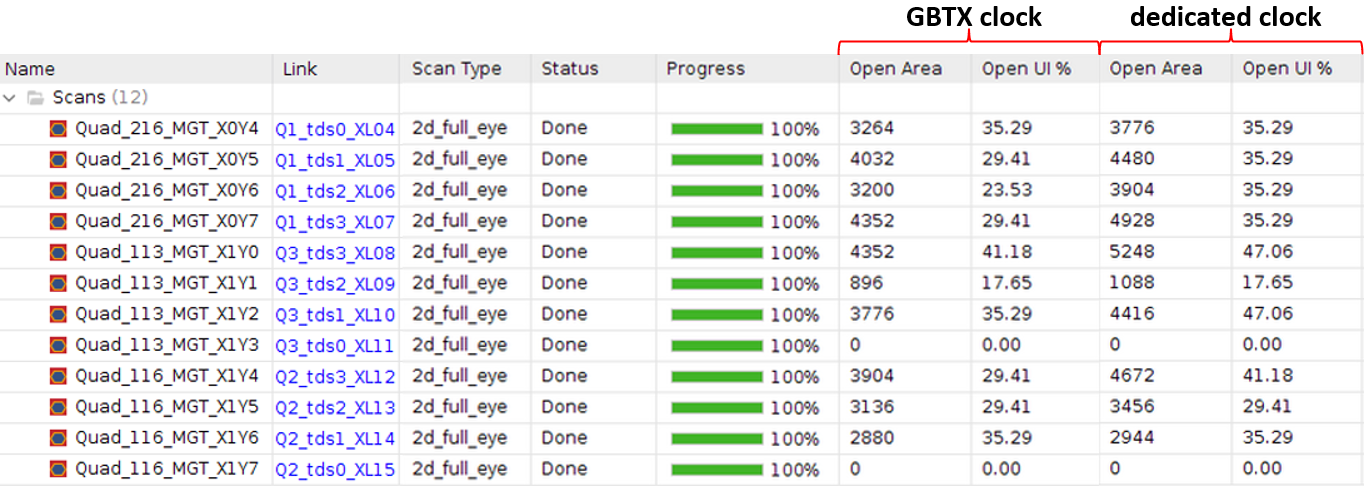}
\caption{\label{fig:eye_comparison} Eye diagram comparison of the $4.8\,\mathrm{Gb/s}$ links for both GBTX and direct clocks using IBERT at the router boards.}
\end{figure}

The eye diagram comparison between the GBTX and direct clock for different links at $4.8\,\mathrm{Gb/s}$ using the SLVS standard is presented in Figure \ref{fig:eye_comparison}. 
The direct clock can increase the open area up to 18\% and the UI up to 33\% more.
    
\begin{figure}[htbp]
\centering 
\includegraphics[width=1\textwidth]{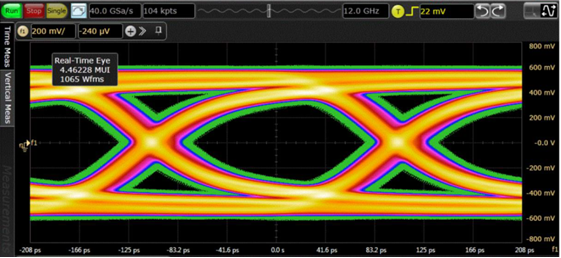}
\caption{\label{fig:eye_osc} The 4.8\,Gb/s serial links were assessed also with the eye diagram by using a high bandwidth oscilloscope.}
\end{figure}

The $4.8\,\mathrm{Gb/s}$ links were also evaluated with the use of a $12\,\mathrm{GHz}$ high performance oscilloscope (DSA91204A by Infiniium). 
The eye diagram of the SLVS signal over $8\,\mathrm{m}$ silver-plated cable, a repeater and $0.5\,\mathrm{m}$ tin-plated cable is presented in Figure\,\ref{fig:eye_osc}.

\section{Conclusions}

A new scheme using COTS components was developed for the distribution of a low-jitter clock to the FPGA transceivers of the NSW sTGC trigger electronics. 
This scheme can achieve a maximum of $1\,\mathrm{ps}$ jitter measured at the output of the Rim-L1DDC board over $110\,\mathrm{m}$ cables (almost double the required length) and the phase noise remains in general bellow the mask provided by Xilinx\textsuperscript{\textregistered}. 
In total 64 clocks can be distributed by the two clock distributor boards to the L1DDC and from there to the 288 trigger boards with the use of fanout chips.
The low jitter clocks with the combination of the repeater chips provide a robust solution for the transmission of the trigger data over long cables using the SLVS standard.
IBERT tests on the $4.8\,\mathrm{Gb/s}$ links over $5\,\mathrm{m}$ twinax cables with a BER of $10^{-14}$ and zero errors, but also eye diagrams on the high speed serial links proved that the new scheme can work reliably. 
Finally, by using the final NSW electronics and infrastructure it was proved that the direct clock can improve the signal quality of the high speed signals compared to the GBTX clock.


\acknowledgments
We acknowledge support of this work by the project “DeTAnet:\ Detector Development and Technologies for High Energy Physics” (MIS\,5029538) which is implemented under the action 
“Reinforcement of the Research and Innovation Infrastructure” funded by the Operational Programme “Competitiveness, 
Entrepreneurship and Innovation” (NSRF\,2014–2020) and co-financed by Greece and the European Union (European Regional Development Fund).



\end{document}